\documentclass[a4paper,fleqn,usenatbib]{mnras}

\usepackage[T1]{fontenc}
\usepackage{ae,aecompl}
\usepackage{graphicx,longtable}
\usepackage{amsmath}
\usepackage{amssymb}
\usepackage{subfig}
\usepackage{soul}
\usepackage{multirow}
\usepackage[dvipsnames]{xcolor}

\newcommand{\hequad}{HE\,0435$-$1223}
\newcommand{\bquad}{B1608$+$656}

\newcommand{\kappaext}{\kappa_{\mathrm{ext}}}
\newcommand{\kappawl}{{\kappa_{\mathrm{wl}}}}
\newcommand{\kappaextwl}{\kappa_{\mathrm{ext}\leftarrow\mathrm{wl}}}
\newcommand{\kappaextnc}{\kappa_{\mathrm{ext}\leftarrow\mathrm{nc}}}
\newcommand{\Lextwl}{L_{\mathrm{ext}\leftarrow\mathrm{wl}}}
\newcommand{\Lextnc}{L_{\mathrm{ext}\leftarrow\mathrm{nc}}}

\newcommand{\data}{{\mathbf{d}}}

\newcommand{\arcsect}{\mathrm{arcsec}}
\newcommand{\kpc}{\mathrm{kpc}}
\newcommand{\kmspc}{${\textrm{km}\, \textrm{s}^{-1} \textrm{Mpc}^{-1}}$}

\usepackage{natbib}

\title[External convergence in the field of \bquad]{H0LiCOW XI. A weak lensing measurement of the external convergence in the field of the lensed quasar \bquad\ using HST and Subaru deep imaging}

\author[O.~Tihhonova et al.]{
O.~Tihhonova,$^{1}$\thanks{E-mail: olga.tihhonova@gmail.com}
F.~Courbin,$^{1}$
D.~Harvey,$^{2,1}$
S.~Hilbert,$^{3,4}$
A.~Peel,$^{1,5}$
C.~E.~Rusu,$^{6, 7, 8}$
\newauthor
C.~D.~Fassnacht,$^{8}$
V.~Bonvin,$^{1}$
P.~J.~Marshall,$^{9}$
G.~Meylan,$^{1}$
D.~Sluse,$^{10}$
\newauthor
S.~H.~Suyu,$^{11,12,13}$
T.~Treu,$^{14, 15}$
K.~C.~Wong$^{16}$
\\
	$^{1}$Institut de Physique, Laboratoire d'Astrophysique, Ecole Polytechnique F{\'e}d{\'e}rale de Lausanne (EPFL),\\
	      Observatoire de Sauverny, CH-1290	Versoix, Switzerland\\
	 $^{2}$ Instituut-Lorentz for Theoretical Physics, Universiteit Leiden, Niels Bohrweg 2, Leiden, 2333 CA, The Netherlands\\     
	$^{3}$Exzellenzcluster Universe, Boltzmannstr. 2, 85748 Garching, Germany \\
	$^{4}$Ludwig-Maximilians-Universit{\"a}t, Universit{\"a}ts-Sternwarte, Scheinerstr. 1, 81679 M{\"u}nchen, Germany \\
	$^{5}$AIM, CEA, CNRS, Universit{\'e} Paris-Saclay, Universit{\'e} Paris Diderot, Sorbonne Paris Cit{\'e}, F-91191 Gif-sur-Yvette, France \\
	$^{6}$Subaru Telescope, 650 N Aohoku Pl, Hilo, HI 96720, USA\\
	$^{7}$Subaru Fellow \\
	$^{8}$Department of Physics, University of California, Davis, CA 95616, USA\\
	$^{9}$Kavli Institute for Particle Astrophysics and Cosmology,
	      Stanford University, 452 Lomita Mall, Stanford, CA 94035, USA\\
	$^{10}$STAR Institute, Quartier Agora - All\'ee du six Ao\^ut, 19c B-4000 Li\`ege, Belgium\\
	$^{11}$Max-Planck-Institut f{\"u}r Astrophysik, Karl-Schwarzschild-Str.~1, 85748 Garching, Germany\\
	$^{12}$Institute of Astronomy and Astrophysics, Academia Sinica, 11F of ASMAB, No.1, Section 4, Roosevelt Road, Taipei 10617, Taiwan\\
	$^{13}$Physik-Department, Technische Universit\"at M\"unchen, James-Franck-Stra\ss{}e~1, 85748 Garching, Germany\\
	$^{14}$Department of Physics and Astronomy, University of California, Los Angeles, CA 90095, USA \\
	$^{15}$Packard Fellow \\
	$^{16}$Kavli IPMU (WPI), UTIAS, The University of Tokyo, Kashiwa, Chiba 277-8583, Japan
}

\date{Accepted XXX. Received YYY; in original form ZZZ}
\pubyear{2019}

\begin{document}
\label{firstpage}
\pagerange{\pageref{firstpage}--\pageref{lastpage}}
\maketitle

\begin{abstract}
	We investigate the environment and line of sight of the H0LiCOW lens \bquad\ using Subaru Suprime-Cam and the Hubble Space Telescope (HST) to perform a weak lensing analysis. We compare three different methods to reconstruct the mass map of the field, i.e. the standard Kaiser-Squires inversion coupled with inpainting and Gaussian or wavelet filtering, and $\tt{Glimpse}$ a method based on sparse regularization of the shear field. We find no substantial difference between the 2D mass reconstructions, but we find that the ground-based data is less sensitive to small-scale structures than the space-based observations. Marginalising over the results obtained with all the reconstruction techniques applied to the two available HST filters F606W and F814W, we estimate the external convergence, $\kappaext$ at the position of \bquad\ is $\kappaext = 0.11^{+0.06}_{-0.04}$, where the error bars corresponds respectively to the 16th and 84th quartiles. This result is compatible with previous estimates using the number-counts technique, suggesting that \bquad\  resides in an over-dense line of sight, but with a completely different technique. Using our mass reconstructions, we also compare the convergence at the position of several groups of galaxies in the field of \bquad\  with the mass measurements using various analytical mass profiles, and find that the weak lensing results favor truncated halo models.
	
	

\end{abstract}


\section{Introduction} \label{sec:intro}

Time delays (TD) in strongly lensed quasars allow one to determine the Hubble constant $H_0$ \citep[e.g.][]{2017MNRAS.468.2590S, 2017MNRAS.465.4914B, 2019MNRAS.484.4726B} independentely of and complementary to other cosmological probes including the Cosmic Microwave Background \citep[CMB;][]{2016A&A...594A..13P}, Baryon Acoustic Oscillations (BAO)+CMB \citep{2017MNRAS.470.2617A}, weak lensing+BAO+Big Bang Nucleosynthesis (BBN) \citep{2018MNRAS.480.3879A}, Cepheids and type Ia supernovae \citep{2012ApJ...758...24F, 2016ApJ...826...56R}, megamasers \citep{2013ApJ...767..154R}, the giant ionized $\mathrm{H}_2$ regions \citep{2018MNRAS.474.1250F}, and standard sirens such as gravitational waves \citep{2017Natur.551...85A}.

The original idea by \citet{Refsdal64} proposes to measure the TDs in gravitationally lensed and photometrical variable sources. When this source is a quasar, the light curves of its lensed images display similar variations, but shifted in time by the TD. This TD is due to 1- to the difference in length between the light paths to its lensed images and 2- to the potential well of the lensing galaxy(ies).  It is proportional to the combination of the three angular diameter distances to the source, to the lens and between the lens and the source. This distance is called the time-delay distance \citep[e.g.][]{schneider2006, 2010ApJ...711..201S}, which is the quantity that TD measure. The time delay distance is then inversely proportional to the Hubble constant, $H_0$, and has a weaker dependence on other cosmological parameters, notably the curvature and the dark energy \citep[e.g.][]{2009ApJ...706...45C}. 

In practice, however, the integrated mass from galaxies and Large Scale Structures (LSS) all the way to the redshift of the lensed quasar also contribute to the lensing potential and can bias the Hubble if not taken into account \citep{2004ApJ...612..660K, 2017ApJ...836..141M, 2016MNRAS.462.1405J}. In gravitational lensing this mass, which can be seen as an external contribution to the potential of the lens, is expressed as a dimensionless surface mass density called external convergence $\kappa_{\mathrm{ext}}$. If $H_0^{\rm model}$ is the value predicted by a lens model that does not incorporate explicitly the line-of-sight (LoS) contribution, then the actual value of $H_0$ is
\begin{equation}
H_0 = (1-\kappaext) \times H_0^{\mathrm{model}}.
\label{kappaext}
\end{equation}
A positive value of $\kappa_{\mathrm{ext}}$ corresponds to a LoS which is denser than the average mass density of the rest of the universe and the true value of $H_0$ is over-estimated by the model. Conversely a negative value of $\kappa_{\mathrm{ext}}$ corresponds to an underdense LoS then the models under-estimate the Hubble constant.

The COSMOGRAIL collaboration \citep[the COSmological MOnitoring of GRAvItational Lenses;][]{2005IAUS..225..297C} measures TDs in lensed quasars using decade-long optical light curves of the quasar images \citep[e.g.][]{2013A&A...553A.121E, 2013A&A...556A..22T, 2017MNRAS.465.4914B}, and has recently adopted a new high-cadence strategy to measure TDs in only 1 observing season \citep{2018A&A...609A..71C, 2018A&A...616A.183B}. 
These TDs are used by the H0LiCOW collaboration \citep[$H_0$ Lenses in COSMOGRAIL's Wellspring;][]{2017MNRAS.468.2590S} to measure the Hubble constant. 

Using deep sharp HST and Adaptive Optics images in combination with dynamics of the main lensing galaxy and spectroscopy of field galaxies \citep[e.g.][]{2017MNRAS.470.4838S, 2017MNRAS.465.4895W, 2017MNRAS.465.4634D}, H0LiCOW has measured $H_0$ to $4\%$ precision \citep{2017MNRAS.465.4914B} using three strongly lensed systems: \bquad\, RXJ1131$-$1231 and \hequad. This was improved to 3\% by adding the fourth system SDSS\,1206$+$4332 \citep{2019MNRAS.484.4726B}. For each of these lenses the effect of the LoS was taken into account using the galaxy number counts technique \citep{2010ApJ...711..201S, 2013ApJ...766...70S, 2017MNRAS.467.4220R, 2019MNRAS.484.4726B} in combination with ray-tracing in the Millennium Simulation \citep[MS,][]{2005Natur.435..629S, 2009A&A...499...31H}. This statistical approach measures the galaxy number density in the vicinity of the lens and compares it to reference fields, hence determining if the LoS is over- or under-dense compared to average. Similar lines of sight and then searched for in the MS. The external convergence calculated for these LoS in the MS are then used to obtain a probability distribution for $\kappaext$. 

In this work, we use a different approach to constrain $\kappa_{\mathrm{ext}}$.
We employ weak gravitational lensing, which is sensitive to both luminous and dark matter, to characterize the LoS to the lensed quasar. In general, the external convergence $\kappa_{\mathrm{ext}}$ to the quasar and the convergence $\kappawl$ inferred from weak lensing mass reconstruction differ, but are statistically dependent. When knowing this dependence, one can thus constrain $\kappaext$ by measuring $\kappawl$ and computing the conditional probability $P(\kappaext|\kappawl)$. We therefore select LoS in the MS with a weak lensing signal similar to measured one. For these LoS we then compute the external convergence to the quasar redshift and use these values to estimate $P(\kappaext|\kappawl)$. This approach was previously applied to the \hequad\ system in \citet{2018MNRAS.477.5657T}, showing no significant LoS contribution for this specific object and agreeing with the galaxy number counts technique. We now apply the method to \bquad, which is the first lens analyzed in the H0LiCOW sample. For this purpose we use deep HST data in addition to ground-based Subaru data. 

While \hequad\ was found to reside in a slightly under-dense environment \citep{2017MNRAS.467.4220R, 2018MNRAS.477.5657T}, somewhat atypical for strong gravitational lenses \citep{2016MNRAS.462.3255C} as they are massive galaxies, \bquad, on the contrary, was found to lie in a crowded field \citep{2011MNRAS.410.2167F}, making this system interesting for our method. While the contribution of the external convergence on the determination of the Hubble constant is marginal in the case of \hequad\ \citep{2017MNRAS.467.4220R, 2018MNRAS.477.5657T}, it gets as high as $10\%$ for \bquad\ if neglected \citep{2010ApJ...711..201S}. This makes a complementary and independent analysis even more important. 

The paper is organized as follows. Section~\ref{sec:wl} reminds the main suite of techniques adopted to reconstruct our convergence maps. Section~\ref{sec:data} presents the ground-based Subaru data and the HST data. In Section~\ref{sec:source_selection} we describe the selection of galaxies used to measure the weak lensing signal and the resulting convergence maps are shown in Section~\ref{sec:convergence_maps}. Section~\ref{sec:joint_convergence_distribution} assesses the impact of the redshift distribution of the galaxy selection. Our results and their comparison with previous works is detailed in Section~\ref{sec:results}. Finally, we conclude in Section~\ref{sec:summary}.  

We adopt a flat $\Lambda$CDM cosmology with $\Omega_{\mathrm{M}} = 0.3$,  $\Omega_{\Lambda} = 0.7$, and $H_0=70$ \kmspc\  when producing mocks of observations. Note that the details of the cosmology chosen here have no significant impact the measured $\kappa_{\mathrm{ext}}$.

\section{Mass mapping methodology}
\label{sec:wl}

The goal of this work is to use weak gravitational lensing to constrain the dimensionless surface mass density, i.e. the convergence $\kappaext$, at the position on the sky of the quadruply imaged quasar \bquad. A detailed account of the weak lensing formalism can be found in \citet{2001PhR...340..291B} and is also explained in \citet{2018MNRAS.477.5657T}. It can be summarized as follows. 

When expressed in units of the critical surface mass density $\Sigma_{\mathrm{crit}}$, the convergence seen by a source at redshift $z_{\mathrm{s}}$ is given by
\begin{equation}
\kappa (\pmb{\theta}, z_{\mathrm{s}}) = \int_{0}^{z_{\mathrm{s}}} \frac{\Sigma (\pmb{\theta}, z)}{\Sigma_{\mathrm{crit}}} \; \mathrm{d}z,
\end{equation}
where $\pmb{\theta}$ is the 2D angular position on the sky, $\Sigma (\pmb{\theta}, z)$ is the surface mass density in the redshift interval ${\mathrm{d}}z$, and the critical surface mass density $\Sigma_{\mathrm{crit}}$ is
\begin{equation}
\Sigma_{\mathrm{crit}} = \frac{c^2}{4 \pi G} \frac{D_{\mathrm{os}}}{D_{\mathrm{od}} D_{\mathrm{ds}}}.
\end{equation}
$D_{\mathrm{od}}$, $D_{\mathrm{ds}}$, $D_{\mathrm{os}}$ are the angular diameter distances between the observer and the deflector, the deflector and the source, and the observer and the source respectively.

The background sources used in weak lensing span a redshift interval with a given distribution $p(z_{\mathrm{s}})$. The cumulative convergence of such a distribution reads
\begin{equation}
\kappa(\pmb{\theta}) = \int_{0}^{z_{\mathrm{max}}} \kappa (\pmb{\theta}, z_{\mathrm{s}}) \; p(z_{\mathrm{s}}) \; \mathrm{d} z_{\mathrm{s}},
\end{equation}
where the integration is carried along the line of sight up to a maximum redshift $z_{\mathrm{max}}$.

\subsection{Kaiser-Squires mass reconstruction and inpainting}
\label{sec:theor_KS}

In the weak lensing regime, foreground lenses only slightly distort the apparent shape of galaxies located further away, i.e. the sources. This small distortion can be measured in practice as a change in ellipticity of the source galaxies. This is known as the complex shear, $\gamma(\pmb{\theta})$,  at any angular position $\pmb{\theta}$ on the plane of the sky. This shear field can be used to reconstruct the underlying convergence field $\kappa(\pmb{\theta})$ as first proposed by \citet[][hereafter KS]{1993ApJ...404..441K}, who solve for the equation
\begin{equation}
\kappa(\pmb{\theta}) = \frac{1}{\pi} \int d^2 \pmb{\theta}' \; \Re [ D^*(\pmb{\theta} - \pmb{\theta}') \gamma(\pmb{\theta}')],
\label{kappa_KS}
\end{equation}
where $D(\pmb{\theta}) = - (\pmb{\theta}_1 - i \pmb{\theta}_2)^{-2}$ is a complex convolution kernel. 

In practice, the shear at any given position on the sky is estimated by measuring the mean complex ellipticity of galaxies in spatial bins, which size will affect the spatial resolution of the mass maps as well as their signal-to-noise. In addition, some source galaxies are masked by bright stars or blended with other galaxies, so that the data used for the reconstruction are incomplete.

For this reason we use the inpainting technique, implemented in the $\tt{FASTLens}$\footnote{\url{http://www.cosmostat.org/software/fastlens/}} package \citep{2009MNRAS.395.1265P}. The core of the technique is to extrapolate the existing data using a sparsity prior. It assumes that there exists a transformation, where the full data (i.e. as if no source galaxy was lost) is sparser than the actual data available for the reconstruction. This permits to increase the binning of the shear map, and therefore the final resolution, so that there is approximately one galaxy per space element. Note, however, that we are not using the inpainted data in the final results, as it is just a statistical extrapolation used to reduce noise. In addition, there are no missing source galaxies at the position of \bquad\ in our Subaru and HST images.

The inpainting technique does not fully remove the noise and the convergence field produced from Eq~\ref{kappa_KS} still needs filtering. Originally, \citet{1993ApJ...404..441K} applied a Gaussian kernel preserving the noise gaussianity, but washing out the small spatial scales in the reconstructed maps. In the present work we use a Gaussian filter alongside the multi-scale entropy filtering \citep[MSE;][]{2006aida.book.....S}, which is a non-linear Bayesian filtering technique based on wavelets. We use the MSE algorithm implemented in the $\tt{MRLens}$\footnote{\url{http://www.cosmostat.org/software/mrlens/}} software \citep{2006A&A...451.1139S}. The noise level is controlled by the Proportion of False Discoveries (PFD) parameter $\alpha$ \citep{2001AJ....122.3492M}, which determines the fraction of the peaks detected as real, but which are in reality due to noise. Setting $\alpha = 0.01$ means that 1\% of the detected peaks are in fact noise \citep[see page 6 of][]{2001AJ....122.3492M}. 

\subsection{Mass reconstruction using $\tt{Glimpse}$}
\label{sec:theor_glimpse}

The KS methods requires to work that the shear field is measured on a regular grid, resulting in loss of spatial resolution and in possible aliasing of signals changing artificially frequency due to the arbitrary binning applied to the data. This can be avoided by using a method that considers the shear field measured at the position of each individual source galaxy and that involves sparse regularisation techniques. 

Such an approach is implemented in the $\tt{Glimpse}$\footnote{\url{http://www.cosmostat.org/software/glimpse}} algorithm \citep{2016A&A...591A...2L}. $\tt{Glimpse}$ uses a multi-scale wavelet-based sparsity prior to solve the regularisation problem, while performing a denoising at the same time. The trade-off between a fidelity term and regularization term is determined with a thresholding parameter $\lambda$, i.e. a Lagrange parameter. Small $\lambda$ values preserve small amplitude coefficients in the wavelet domain, with the risk that they could potentially be due to noise. Larger values of $\lambda$ leave only the most significant peaks while potentially removing small amplitude signals. 

$\tt{Glimpse}$ can incorporate higher-order distortions such as flexion together with the shear to recover the high resolution convergence maps. However, in this work we only use shear data. $\tt{Glimpse}$ is well suited for recovering cluster information \citep{2017ApJ...847...23P}, and was shown to outperform the standard Kaiser-Squires method with Gaussian filtering in both the convergence field reconstruction and the peak statistics when applied to the Dark Energy Survey Science Verification data \citep{2018MNRAS.479.2871J}.

\section{Observations and Data Reduction}
\label{sec:data}
\begin{figure*}
	\centering
	\captionsetup[subfigure]{labelformat=empty}
	\subfloat{
		\includegraphics[width=1.37\columnwidth]{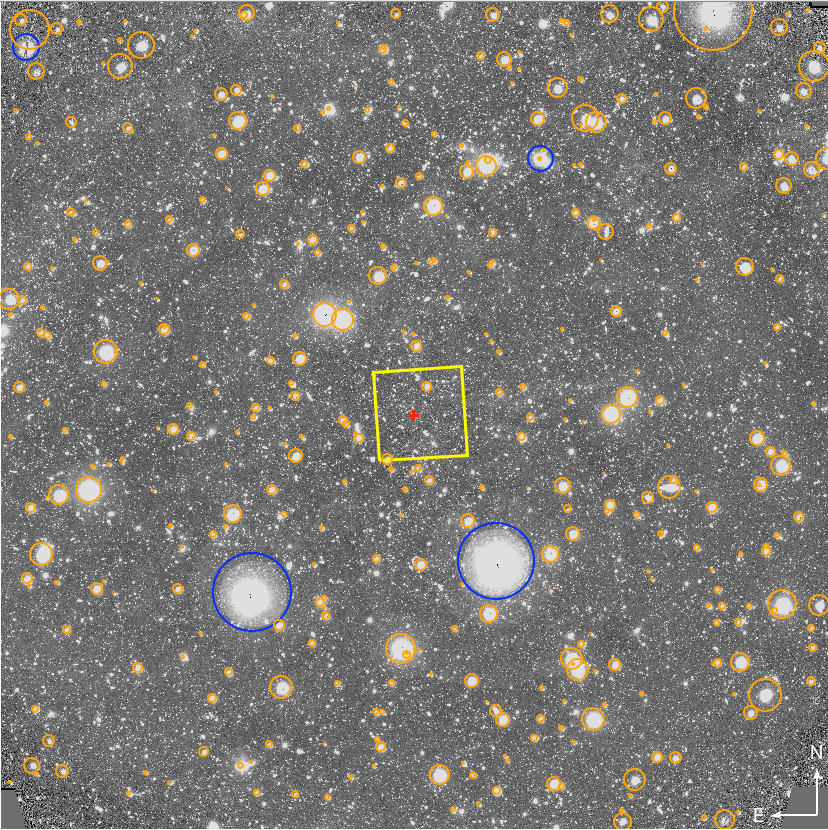}
	}
	\subfloat{
		\includegraphics[width=0.675\columnwidth]{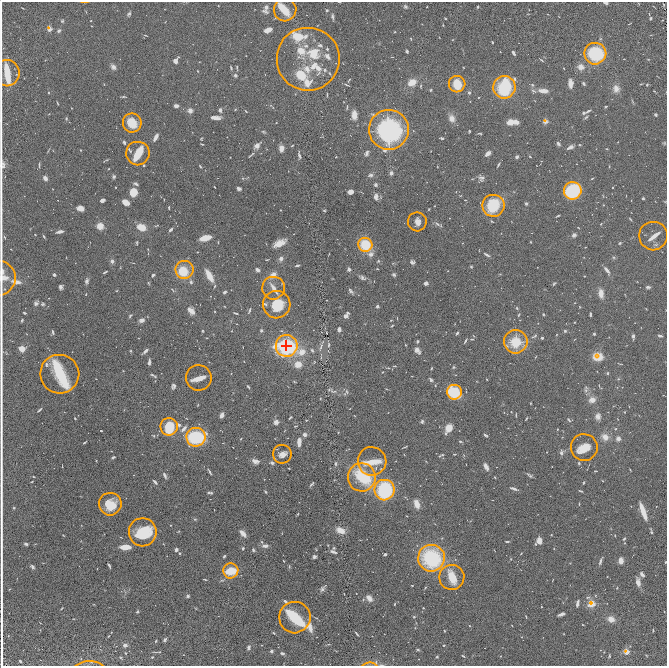}
	}
	\vspace{-12cm}
	
	\hspace{11.7cm}
	\subfloat{
		\includegraphics[width=0.675\columnwidth]{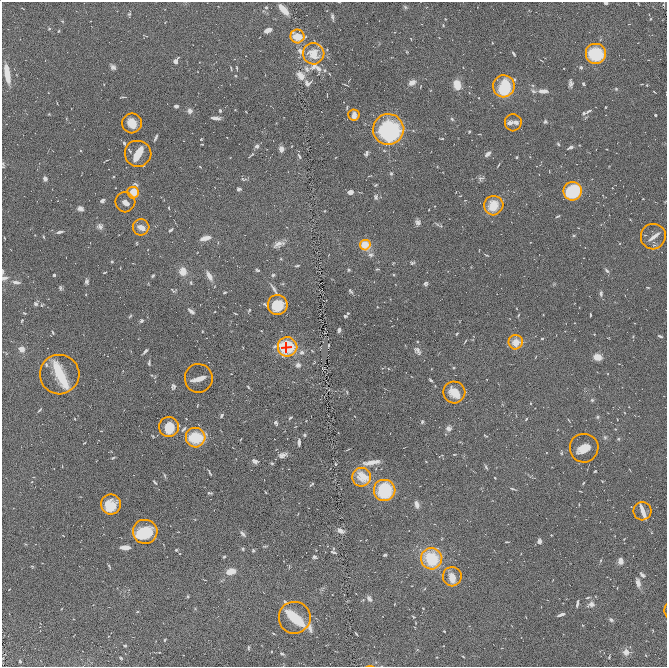}
	}
	\vspace{6cm}
	\caption{The left panel shows the Subaru Suprime-Cam $r$-band $0.5^{\circ} \times 0.5^{\circ}$ field with \bquad\ located in the middle. North is up and East is left. The yellow square in the center delimits the $3.21\arcmin \times 3.21 \arcmin$ extent of the HST data, shown on the right panel with the F606W filter on the top and the F814W filter in the bottom. The red cross shows the position of \bquad. The orange and blue circles indicate the automatic and manual masks for the saturated objects. The final galaxy density of the sources adopted for the weak lensing analysis is 14~gals/arcmin$^2$ and 220~gals/arcmin$^2$ for the ground and the space data respectively.}
	\label{fig:1608}
\end{figure*}

The quadruply imaged quasar \bquad\ was discovered in the Cosmic Lens All-Sky Survey \citep[CLASS,][]{1995ApJ...447L...5M, 2003MNRAS.341....1M, 2003MNRAS.341...13B}. The redshift of the source is $z_{\mathrm{s}} = 1.394$ \citep{1996ApJ...460L.103F}, and the redshift of the two merging lens galaxies is $z_{\mathrm{l}} = 0.6304$ \citep{1995ApJ...447L...5M}, both being part of a small group with $\approx10$ confirmed members \citep{2006ApJ...642...30F}. We use ground-based observations with the Subaru telescope as well as imaging with the Hubble Space Telescope (HST) to estimate the convergence field around the gravitational lens system \bquad\ and evaluated the impact on the measured Hubble constant from the time delays in this system.

\subsection{Subaru Suprime-Cam imaging}

We use multi-band, wide-field imaging observations around \bquad\ at ultraviolet and optical wavelengths. The ground-based data is obtained with the Subaru Telescope (PI. C. Fassnacht) and the Canada-France-Hawaii Telescope (CFHT; PI. S. Suyu, run ID 14AT01), as detailed in Table~\ref{tab:observations}. The CFHT MegaCam data is taken from the CFHT archive pre-reduced and photometrically calibrated with standard techniques. The Subaru Suprime-Cam data is first reduced with SDFRED1 \citep{2002AJ....123...66Y, 2004ApJ...611..660O}, and then processed with Scamp \citep{2006ASPC..351..112B} to achieve a consistent astrometric and photometric calibration, matching the CFHT data. Finally, both datasets are re-sampled with Swarp \citep{2002ASPC..281..228B} to a $0.2\arcsec$ pixel scale, corresponding to the (larger) Suprime-Cam native pixel scale.  We ca\-li\-brate the Suprime-Cam data using observations of a Sloan Digital Sky Survey star field, taken during the same night. Finally, we correct for galactic and atmospheric extinction following \citet{2011ApJ...737..103S} and \citet{2012yCat..35490008B}, respectively.

\begin{table*}
	\centering
		\begin{tabular}{@{}lllllll@{}}
			\hline 
			Telescope/Instrument & Pixel scale [$\arcsec$] & Filter & Exposure [sec] & Airmass & Seeing [$\arcsec$] & Observation date \\ 
			\hline
			CFHT/MegaCam & 0.187 & $u$ & $50 \times 494$ & $1.4 - 2.0$ & $\sim 0.8$ & 2014 Apr. 26 - Jun. 30 \\
			Subaru/Suprime-Cam & 0.200 & $g$ & $5 \times 120$ & $\sim1.6$ & $\sim 0.75$ & 2014 Mar. 1 \\
			Subaru/Suprime-Cam & 0.200 & $r$ & $16 \times 300$ & $\sim1.5$ & $\sim 0.75$ & 2014 Mar. 1 \\
			Subaru/Suprime-Cam & 0.200 & $i$ & $5 \times 120$ & $\sim1.4$ & $\sim 0.95$ & 2014 Mar. 1 \\
			\hline
		\end{tabular}
		\caption{Summary of the ground-based observations with Subaru/Suprime-Cam and CFHT/MegaCam.}
		\label{tab:observations}
\end{table*}

To measure the shape of source galaxies, we use the deep and sharp Subaru Suprime-Cam $r$-band image. The field of view is 34\arcmin $\times$ 27\arcmin, as shown in Fig.~\ref{fig:1608}, and the pixel scale of $0.2 \arcsec$. Before doing any shape measurement, we subtract the sky background using the $\tt{mr\_background}$ algorithm, which is part of the Multiresolution Analysis Software\footnote{\url{http://www.multiresolutions.com/mr/}}. The choice of the parameters is done in the same way as in  \citep[][]{2018MNRAS.477.5657T}. We also mask all saturated objects detected with the $\tt{SExtractor}$ software\footnote{\url{http://www.astromatic.net/software/sextractor}} \citep{1996A&AS..117..393B}. In doing this, we set the radius of the circular masks to be $r = 2\times \mathtt{FWHM\_IMAGE}$, where $\mathtt{FWHM\_IMAGE}$ is the measured Full Width at Half Maximum. We also manually mask 4 bright stars not detected by $\tt{SExtractor}$ (see Fig.~\ref{fig:1608}). The resulting background-subtracted and masked image in shown Fig.~\ref{fig:1608}.

\subsection{HST imaging}

We use deep HST data of \bquad\ observed with the ACS/WFC camera in the F606W and F814W filters in August 2004 (Program ID 10158; PI: C.~D.~Fassnacht). The data are described in more detail in \citet{2009ApJ...691..277S}, with the summary of the exposures given in their Table~2. 

We remove the effect of Charge Transfer Efficiency introduced during the readout following \citet{CTI,CTI2,CTI3}, before calibrating each individual exposure using the publicly available CALACS package. We then combine them using the $\tt{astrodrizzle}$ package with the standard square kernel \citep{CTI4,astrodrizzle}. 

The final drizzled images have a pixel scale of $0.03 \arcsec$ and a $\mathtt{pixfrac}$ of 0.8, considered as the optimal values for accurate weak lensing measurements \citep{drizzlepars}. We also produce distortion-free maps of each individual exposure in order to carry out measurements of the point spread function (PSF). As our mass-mapping algorithms require squared images, we make $3.21\arcmin \times 3.21 \arcmin$ cutouts with \bquad\ located about $15\arcsec$ away from the center as shown in Fig.~\ref{fig:1608}.

\section{Selection of the source galaxies}
\label{sec:source_selection}

\begin{figure}
	\centering
	\includegraphics[width=1\columnwidth]{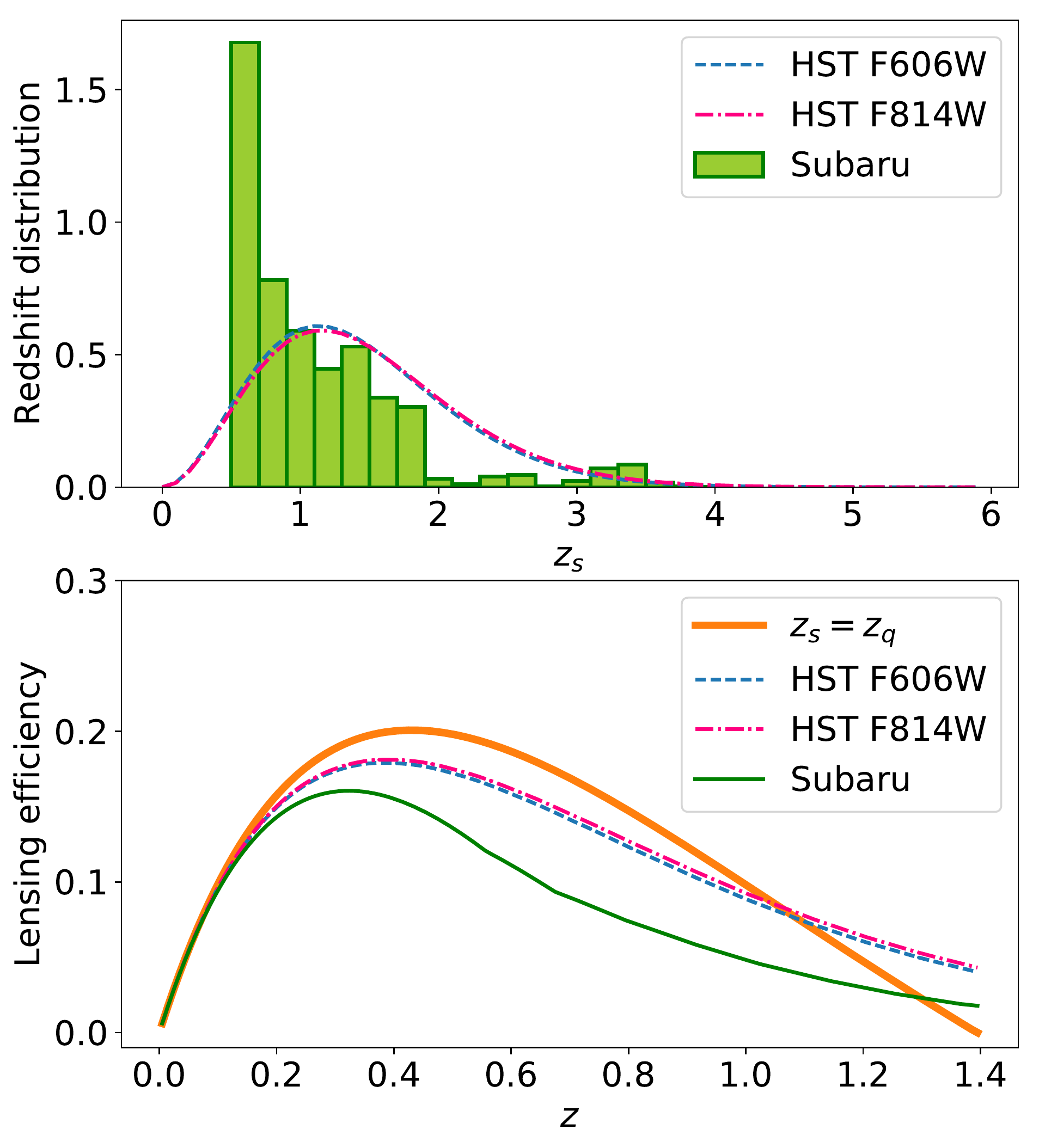}
	\caption{The upper panel shows the final redshift distributions of the selected source galaxies. The green histogram represents the Subaru $r$-band data, while the blue dashed and pink dash-dotted lines show the parametrized distributions for the HST F606W and the HST F814W filters respectively. The lower panel depicts the cumulative lensing efficiency kernels for these galaxies, as compared to the `ideal' kernel for the galaxies as if they are all at the redshift of the quasar (thick orange line).}
	\label{fig:kernel}
\end{figure}

To measure a weak lensing convergence $\kappawl$ close to the external convergence $\kappaext$ at the position of the lensed quasar, we would ideally need a screen of source galaxies at the redshift of the quasar, i.e. $z_{\mathrm{q}}=1.394$, to measure the shear on. Then all intervening matter would have exactly the same lensing efficiency for both $\kappaext$ and $\kappawl$.
In practice, however, selecting only galaxies with the same redshift as the quasar drastically reduces the surface number density of useful sources in terms of ellipticity measurements. We thus instead select a redshift range for which the cumulative lensing efficiency kernel \citep[see Fig. 2 of][]{2018MNRAS.477.5657T} is as close as possible to the `ideal' one where all source galaxies would be at $z_{\mathrm{s}}=z_{\mathrm{q}}$ while leaving enough galaxies for the weak lensing analysis. The mismatch between the 'ideal' lensing  efficiency kernel and the actual one will be accounted for 
when estimating $P(\kappaext|\kappawl)$ using ray-tracing through the Millennium Simulation (MS), as described in Sect.~\ref{sec:joint_convergence_distribution}.

As explained in \citet{2018MNRAS.477.5657T}, the lensing efficiency kernel for any source with redshift $z_{\mathrm{s}}$ and lens (deflector) at redshift  $z_{\mathrm{d}}$ is given by 
\begin{equation}
\label{eff_kernel}
g(z_{\mathrm{d }}, z_{\mathrm{s}}) = \frac{4 \pi G}{c^2} \frac{D_{\mathrm{od}} D_{\mathrm{ds}}}{D_{\mathrm{os}}} \; H(z_{\mathrm{d}}-z_{\mathrm{s}}) = \frac{H(z_{\mathrm{d}}-z_{\mathrm{s}})}{\Sigma_{\mathrm{crit}}},
\end{equation}
where $H(z-z_{\mathrm{s}})$ is the Heaviside step function. The cumulative lensing efficiency kernel is defined as follows:
\begin{equation}
\mathcal{G}(z_{\mathrm{d}}) = \frac{\sum\limits_{i} \; g(z_{\mathrm{d}}, z_{\mathrm{s}}^i) \; n(z_{\mathrm{s}}^i)}{\sum\limits_{i} n(z_{\mathrm{s}}^i)},
\end{equation}
where $n(z_{\mathrm{s}}^i)$ is the number of source galaxies per redshift bin~$i$, and $z_{\mathrm{s}}^i$ is the redshift at the center of the bin 
and where the summation is over all redshift bins of the source galaxies. The lower panel of Fig.~\ref{fig:kernel} illustrates and compares the 'ideal' kernel, i.e. as if all sources where at the redshift of the quasar, with kernels where the redshift distribution of sources is different.

\subsection{The Subaru data}

For the ground-based Subaru data we perform PSF measurement, source detection, galaxy-star classification and photometric redshift estimation using the technique described in \citet{2017MNRAS.467.4220R}. In order to calibrate the photometric redshifts, we use a spectroscopic catalog composed of 134 objects within a radius of $\theta \lesssim4\arcmin$ around the lens, of which we use only the 113 galaxies with the most reliable spectroscopic redshifts. The spectroscopic redshifts were obtained with the Magellan 6.5m telescope \citep{2006ApJ...641..169M, 2015ApJS..219...29M}, and the Keck 10m Telescope (PI: Fassnacht). We discard all the galaxies with photometric redshift uncertainties larger than $\sigma_{\mathrm{z}}= 0.3 \times (1 + z)$, where $\sigma_{\mathrm{z}}$  corresponds to the 68\%-level error. 

We adopt a redshift range of $0.5 < z_{\mathrm{s}} < 4.0$, which gives as a result a total of 12219 source galaxies, i.e. a surface number density of 14~gals/arcmin$^2$. The corresponding cumulative lensing efficiency kernel is shown in Fig.~\ref{fig:kernel}. This kernel is optimal in the sense that it allows to keep a high number density of source galaxies and it is close to the 'ideal' kernel shown in orange in Fig.~\ref{fig:kernel}.

\subsection{The HST data}
\label{HST}

With two HST filters, F606W and F814W, we cannot infer photometric redshifts for the source galaxies. Though several of the HST galaxies are identified in the Subaru field and thus have the photometric redshift information, they represent only $\approx 1/3$ of the population, and their redshift distribution is not representative, as the Subaru data is shallower than that of the HST. We therefore adopt the empirical parametrization of redshift as a function of apparent magnitude $m$ by \citet{1993MNRAS.265..145B}:
\begin{equation}
\begin{cases}
\frac{{\mathrm{d}}N (z, m)}{{\mathrm{d}}z} \propto n_0(m) \frac{z^2}{z_{\mathrm{c}}^3} \exp \left[ - \left( \frac{z}{z_c(m)}  \right) ^{3/2} \right] \\
z_{\mathrm{c}} (m) = z_{\mathrm{m}}/1.412, 
\end{cases}
\label{n_z}
\end{equation}
where $n_0$ is the galaxy density and $z_{\mathrm{m}}$ is the median redshift as a function of magnitude. Note that we randomly sample redshifts from the distribution in the lensing simulations, hence taking the sampling variance into account by construction. 
To derive $z_{\mathrm{m}} (m)$ we use the GOODS-South field \citep[GOODS-S,][]{2004ApJ...600L..93G}, which has a very similar depth in the F606W and the F814W filters compared to our observations \citep{2011ApJS..197...36K, 2011ApJS..197...35G}. The photometric redshifts in GOODS-S are estimated by \citet{2014ApJS..214...24S} using 26 broad bands and 14 medium bands with the EAZY code \citep{2008ApJ...686.1503B}. 

We split our data in magnitude bins of $\Delta m = 0.25$ and estimate $z_{\mathrm{m}}$ for both F606W and F814W. We find perfect agreement with \citet{2007ApJS..172..219L}, who carried out the same analysis in the COSMOS HST ACS field for the F814W filter \citep{2007ApJS..172...38S, 2007ApJS..172..196K}. 

For the same magnitude bins we estimate $n_0$ using the data for \bquad. We then calculate the redshift distribution following Eq.~\ref{n_z} for each magnitude bin, and sum them to get the final distribution, presented on the upper panel of Fig.~\ref{fig:kernel}. We then apply magnitude cuts of $24 < m < 29$ for both the F606W and the F814W filters. 
This leaves 2734 galaxies or 223~gals/arcmin$^2$,  and 2694 galaxies or 220~gals/arcmin$^2$ for the F606W and the F814W filters respectively. The resulting kernels are shown on the lower panel of Fig.~\ref{fig:kernel}. We also check the color-magnitude diagram to make sure we don't include any members of the group where the lens galaxies reside.

\section{Reconstruction of the weak lensing convergence field}
\label{sec:convergence_maps}

\begin{figure*}
	\centering
	\captionsetup[subfigure]{labelformat=empty}
	\subfloat{
		\includegraphics[width=0.7\textwidth]{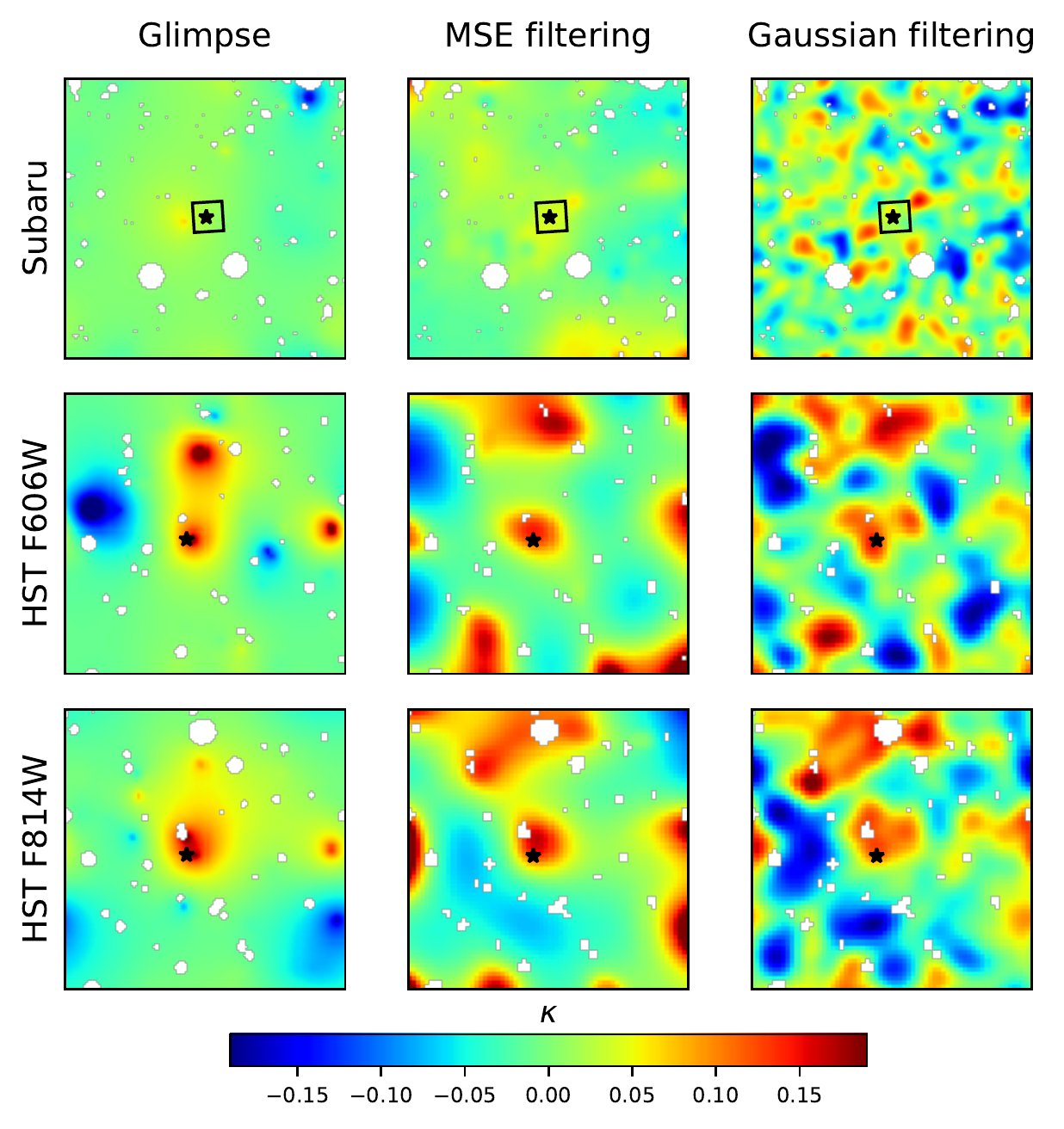}}
	\caption{Weak lensing convergence maps for the \bquad\ field, with the lensed system indicated by a star. The upper row shows the convergence from the Subaru $r$-band data with $0.5^{\circ} \times 0.5^{\circ}$ field of view, the middle and the bottom rows show the convergence from the HST F606W and the HST F814W filter data respectively. The spatial extent of the HST field is $3.21\arcmin \times 3.21 \arcmin$, and is shown by a black square inside the Subaru field. The left column corresponds to the Glimpse reconstruction, the middle column to the MSE filtering. The right column shows the Gaussian filtering with $\theta_{\mathrm{G}} = 0.75 \arcmin$ and $\theta_{\mathrm{G}} = 0.15 \arcmin$ for the Subaru and HST data, respectively. The white regions correspond to masked stars and bright foreground objects. The color scale represents the values of the weak lensing convergence $\kappawl$.}
	\label{fig:kappa}
\end{figure*}

For the Subaru data, we use the $\tt{KSB+}$ software \citep{2006MNRAS.368.1323H} to measure the shapes of the selected source galaxies and obtain the shear estimator $\gamma$. For the HST images we use an adapted version of RRG \citep{RRG}, where we first measure the quadrupole and the fourth order moments of the stars and compare them to TinyTim PSF models \citep{tinytim} to obtain a PSF estimate on the individual exposures. We then rotate these to a common frame and combine then to derive the final PSF. This method naturally accounts for rotations, chip boundaries and the for the varying effective exposure time across the image \citep{Harvey15}. 

We adopt two different methods to reconstruct the convergence map from the shear field: the KS inversion, applied to the inpainted shear field implemented in $\tt{FASTLens}$ and $\tt{MRLens}$, and the sparsity-based reconstruction with $\tt{Glimpse}$. While $\tt{Glimpse}$ already incorporates noise filtering, we adopt a Gaussian and wavelet filtering for the KS reconstruction.

We spatially bin the original data in different ways, so that the final spaxel scales in the convergence fields are $14 \arcsec$ on-a-side for the Subaru data. The HST data processed with $\tt{MRLens}$ have a spaxel size of $3 \arcsec$ and of $1.5 \arcsec$ for the $\tt{Glimpse}$ processing. The spaxel scales are chosen to obtain the finest resolution possible without being entirely dominated by the shot noise. Both mass-reconstruction approaches are iterative, with the number of iterations fixed to 300. $\tt{MRLens}$ and $\tt{Glimpse}$ use a wavelet decomposition, and the original images are decomposed into 5 and 4 wavelet scales for the Subaru and the HST data, respectively. 

To filter the noise in $\tt{MRLens}$, we choose a Gaussian kernels with angular sizes of $\theta_{\mathrm{G}} = 0.75 \arcmin$ and $\theta_{\mathrm{G}} = 0.15 \arcmin$ for the Subaru and the HST data respectively. As shown in \citet[][]{2000MNRAS.313..524V} mass maps reconstruction follow a Gaussian statistics. As a consequence, the signal to noise goes roughly as $\sqrt(N)$, where N is the number of galaxies encompassed by the Kernel. For the Subaru data, the kernel encompasses on average 25 galaxies, and for the HST data the (smaller) kernel encompasses 15 galaxies, corresponding to a signal-to-noise per spatial resolution element of respectively SNR(Subaru)$\sim 5$ and SNR(HST)$\sim 4$. For wavelet filtering in $\tt{MRLens}$ we fix $\alpha = 0.01$, i.e. a false detection rate of 1\%, as in \citet{2018MNRAS.477.5657T} 

For $\tt{Glimpse}$, we choose $\lambda = 4$ (see Section~\ref{sec:theor_glimpse}). There is no direct theoretical relation between this Lagrange parameter and the signal-to-noise of the detections in the $\tt{Glimpse}$ weak lensing map. However, it has been shown empirically on real data that $\lambda=3$ corresponds roughly to a signal-to-noise of 3 and that $\lambda=4$ ensures that all noise peaks are removed while not washing out any real signal. Previous applications exploring this include the $\tt{Glimpse}$ mass map of the lensing cluster Abell 520 \citep{2017ApJ...847...23P} and the wide field mass reconstruction of the Dark Energy Survey Science Verification data \citep{2018MNRAS.479.2871J}.

In order to assess the statistical errors empirically both for the $\tt{MRLens}$ and $\tt{Glimpse}$ methods, we use the Jackknife approach. In doing so, we divide the field of view into $n_{\mathrm{JK}}$ equal regions using the $\tt{kmeans}$ clustering package\footnote{\url{https://github.com/esheldon/kmeans\_radec}}, where we adopt $n_{\mathrm{JK}}=300$ for the Subaru data and $n_{\mathrm{JK}}=250$ for the HST data.  The final convergence maps for the Subaru data, the HST F606W and F814W data with the Gaussian filtering, the wavelet filtering and the $\tt{Glimpse}$ reconstructions are shown in Figure~\ref{fig:kappa}. All maps are visually consistent, with a clear indication of an over-density at the position of \bquad. The Subaru data shows less small-scale structures than the HST data due to the lower spatial resolution and the larger-scale smoothing.

\section{Estimating the Joint Convergence Distribution}
\label{sec:joint_convergence_distribution}

\begin{figure*}
	\centering
	\captionsetup[subfigure]{labelformat=empty}
	\subfloat{
		\includegraphics[width=0.7\textwidth]{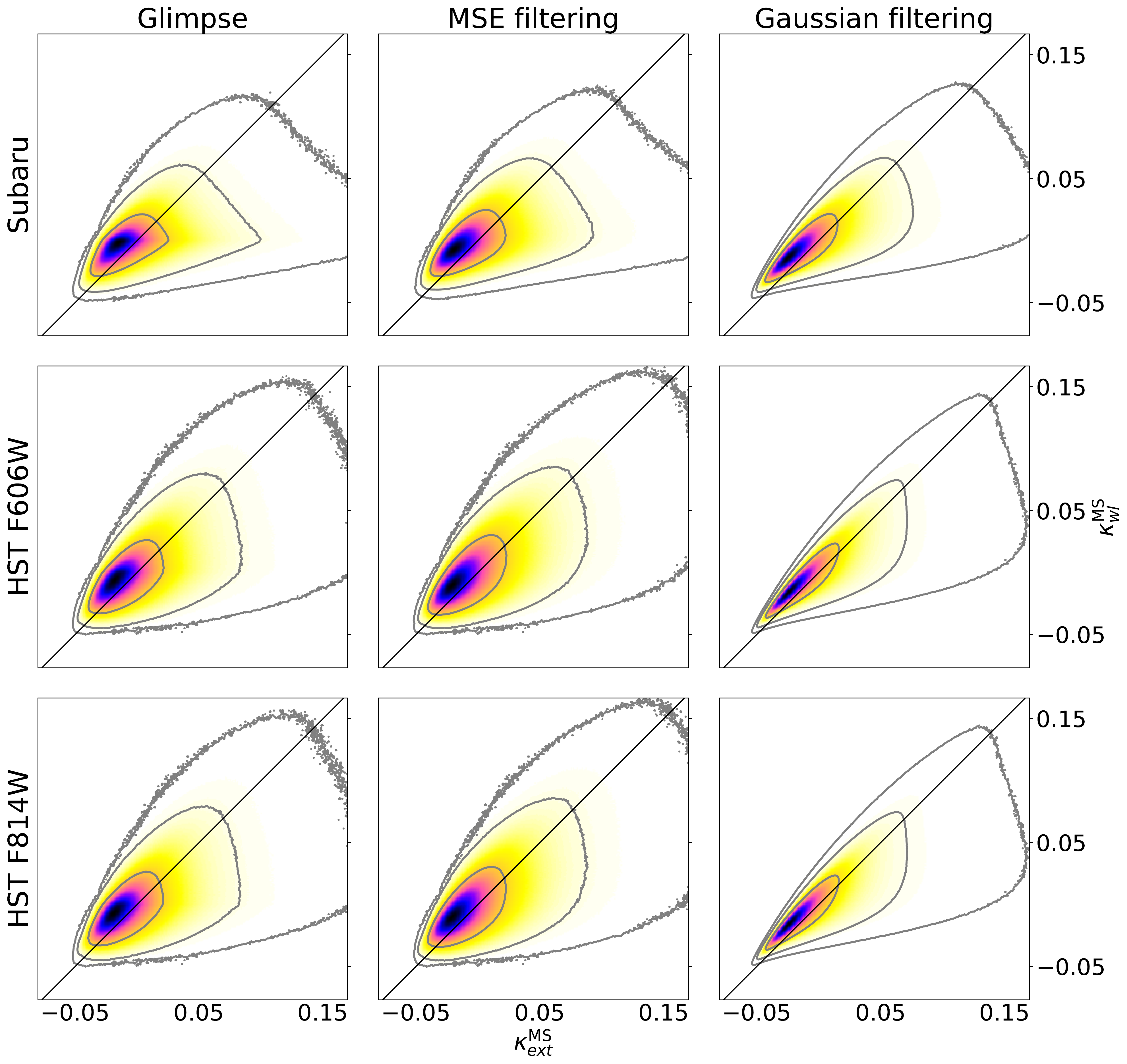}}
	\caption{The joint distribution of the external convergence $\kappaext$ and the weak lensing convergence $\kappawl$, as estimated from the Millennium simulation. Darker colors correspond to higher joint probability density $P(\kappaext, \kappawl)$. The gray contours enclose the $1\sigma$, $2\sigma$ and $3\sigma$ regions. The black diagonal line indicates perfect correlation. The rows from top to bottom depict the Subaru $r$-band, the HST F606W and the HST F814W data. The columns show the reconstructions using the Glimpse, the MSE and the Gaussian algorithms. The Gaussian kernel sizes are $\theta_{\mathrm{G}} = 0.75 \arcmin$ and $\theta_{\mathrm{G}} = 0.15 \arcmin$ for the Subaru and the HST data respectively.}
\label{fig:kernel_distribution}
\end{figure*}

The convergence $\kappawl$ obtained from the weak lensing reconstruction and the external convergence $\kappaext$ the quasar light experiences due to LOS structures differ both statistically and systematically due to several reasons. The source galaxies selected for the weak lensing reconstruction span a redshift distribution, resulting in a lensing efficiency kernel which is different from the ideal one at the redshift of the quasar, as shown in Fig.~\ref{fig:kernel}. Furthermore, the convergence from the reconstruction is smoothed on scales larger than the extent of the strong lens image system, which is the relevant spatial scale for $\kappaext$.  Moreover, the convergence from the reconstruction is subject to noise due to the redshift uncertainties and the intrinsic ellipticities of the source galaxies.

For these reasons, the weak lensing convergence $\kappawl$ cannot be used directly as an estimate of the external convergence $\kappaext$. However, since they share most of the LoS structures as a common cause, they are correlated. Measuring the weak lensing convergence thus still provides very valuable information on the external convergence. The appropriate way to express this information in the framework of Bayesian model parameter estimation employed in H0LiCOW is via $P(\kappaext|\kappawl)$, i.e. the posterior probability distribution of $\kappaext$ given a measurement of $\kappawl$.

The conditional probability distribution $P(\kappaext|\kappawl)$ can be directly computed from the joint distribution $P(\kappaext|\kappawl)$. Following the approach in \citet{2018MNRAS.477.5657T}, we use ray-tracing through the Millennium Simulation \citep[MS,][]{2005Natur.435..629S} populated with galaxies by \citet{2007MNRAS.375....2D} to estimate the joint probability distribution $P(\kappaext,\kappawl)$. Employing multiple-lens-plane ray-tracing \citep{2008MNRAS.386.1845H, 2009A&A...499...31H}, we create 1024 simulated fields of shear and convergence as function of angular position and redshift sampled on grids of $1024^2$ LoS. To reach good sampling of the joint distribution, we do not restrict the analysis to the center of the simulated fields, but instead consider all LoS sufficiently far from the field boundary, such that the weak lensing reconstruction is not affected by boundary effects. For each such LoS, we then record (i) the convergence $\kappaext$ to the quasar redshift, and (ii) the convergence $\kappawl$ obtained from a weak lensing reconstruction performed on the imaging data.

The external convergence $\kappaext$ is directly taken from the convergence to the quasar redshift output by the lensing simulation. The simulation has a spatial resolution of $\sim10\,\kpc$ comoving, which corresponds to an angular resolution of a few $\arcsect$ for the convergence. This resolution is sufficient to capture the angular scales on which LoS structures can act as a mass sheet for the strong lens system. Variations of the convergence on smaller scales not captured because of the limited resolution of the simulation do not act as a mass sheet but instead induce image distortions and require explicit modelling as matter structures.

The weak lensing convergence $\kappawl$ is computed in the following way. We place source galaxies in the simulated fields at the same angular positions as the galaxies in the real Subaru and HST catalogs. For the simulated Subaru catalogs, we set the source galaxy redshifts $z = z_{\mathrm{phot}}$, as we have the photometric redshifts $z_{\mathrm{phot}}$ for all the galaxies in the field. In the HST catalogs, however, only $\approx 1/3$ of the galaxies have $z_{\mathrm{phot}}$. For the simulated HST data, we therefore divide the rest of the population in magnitude bins of $\Delta m = 0.25$ and randomly assign the redshifts $z_{\mathrm{distr}}$ from the distribution in Eq.~\eqref{n_z} per magnitude bin. For statistical purposes, we create 250 different redshift realisations. Thus, depending on the galaxy, we set either $z = z_{\mathrm{phot}}$ or $z = z_{\mathrm{distr}}^i$, where $i = 1, ..., 250$. 

For each of the 1024 MS fields we generate 100 noisy realisations of the source galaxy shear catalog by adding shape noise to the  shear obtained from the simulation at the angular positions and redshifts of the simulated source galaxies. The shape noise is drawn  from a normal distribution with $\sigma_{\gamma} = 0.28$ and $\sigma_{\gamma} = 0.25$ for the Subaru and the HST respectively, as estimated from the data. We then apply our mass reconstruction and filtering algorithms to obtain simulated weak lensing convergence maps $\kappawl$.

The joint analysis of the simulated convergence maps for $\kappaext$ and $\kappawl$ yield $P(\kappaext, \kappawl)$ for all filtering techniques, as shown in Fig.~\ref{fig:kernel_distribution}. After computing $P(\kappaext | \kappawl)$ from that, we  infer the posterior probability distribution of the convergence $P(\kappaext | \data)$ along the line of sight to \bquad\ given the actual shear data $\data$:
 \begin{equation}
 P(\kappaext | \data) = \int \mathrm{d} \kappawl \,\, P(\kappaext | \kappawl) \times P(\kappawl | \data).
 \end{equation}
As previously discussed, statistical sources of noise, such as shape noise, the redshift uncertainty, or the random source galaxy positions are fully accounted for by construction of the method, i.e. by comparing real observations to simulations.

\section{The line of sight convergence}
\label{sec:results}

\begin{figure*}
	\centering
	\captionsetup[subfigure]{labelformat=empty}
	\subfloat{
		\includegraphics[width=0.7\textwidth]{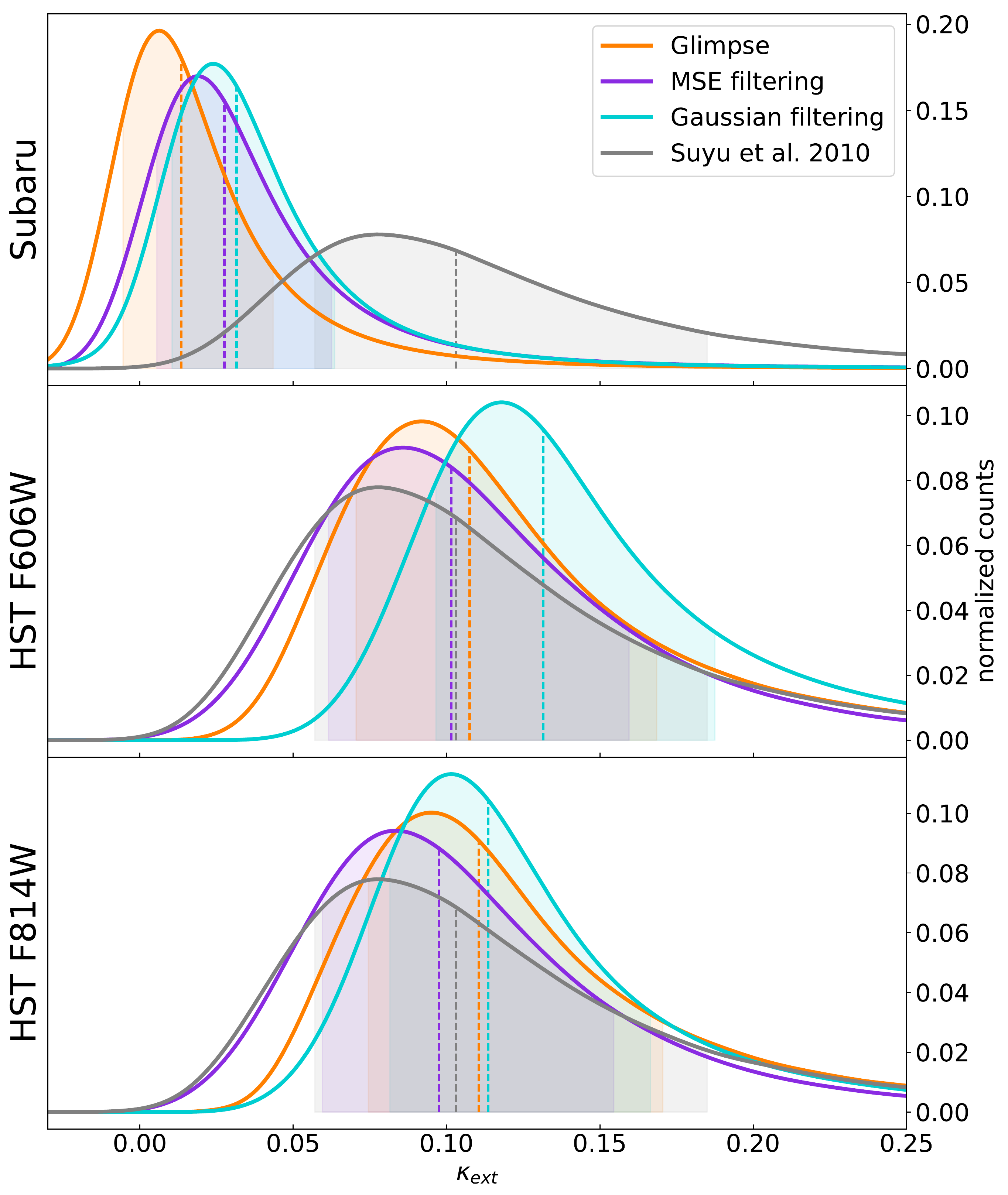}}
	\caption{The PDFs of the external convergence at the position of \bquad\ inferred from the data from Subaru $r$-band, and the HST filters F606W and F814W. The orange distribution corresponds to the Glimpse reconstruction and filtering of the data, the purple to the MSE filtering, and the blue to the Gaussian filtering with $\theta_{\mathrm{G}} = 0.75 \arcmin$ and $\theta_{\mathrm{G}} = 0.15 \arcmin$ for the Subaru and HST data respectively. The gray distribution shows the result from \citet{2010ApJ...711..201S} from the number counts technique. Dashed lines show the values at the $50\%$ percentile. Shaded regions indicate the 16th and 84th percentile.}
	\label{fig:final_result}
\end{figure*}

\begin{table}
	\begin{center}
		\begin{tabular}{ c c c c c c}
			\hline
			Data & Method & $\kappa_{\mathrm{ext}}$ & $\sigma_{\kappa}^{-}$ &  $\sigma_{\kappa}^{+}$ & F \\
			\hline\hline
			\multirow{3}{*}{Subaru} & Glimpse & 0.01 & 0.02 & 0.03 & 2.9 \\
			& MSE & 0.03 & 0.02 & 0.04  & 4.6 \\
			& G$0.75\arcmin$ & 0.03 & 0.02 & 0.03 & 5.1 \\
			\hline
			\multirow{3}{*}{F606W} & Glimpse & 0.11 & 0.04 & 0.06 & 8.7 \\
			& MSE & 0.10 & 0.04 & 0.06  & 8.6 \\
			& G$0.15\arcmin$ & 0.13 & 0.03 & 0.06 & 7.3 \\
			\hline
			\multirow{3}{*}{F814W} & Glimpse & 0.11 & 0.04 & 0.06 & 8.6 \\
			& MSE & 0.10 & 0.04 & 0.06  & 8.8 \\
			& G$0.15\arcmin$ & 0.11 & 0.03 & 0.05 & 8.6 \\
			\hline
			\multicolumn{2} {c} {Suyu et al. 2010} & 0.10 & 0.05 & 0.08 & $-$\\
			\hline
		\end{tabular}
	\end{center}
	\caption{Estimates of $\kappaext$ at the position of \bquad\ using data from the Subaru $r$-band, and the HST filters F606W and F814W. We apply different mass reconstruction and noise filtering methods: Glimpse, MSE filtering, Gaussian filtering with $\theta_{\mathrm{G}} = 0.75\arcmin$ and $\theta_{\mathrm{G}} = 0.15\arcmin$ for the Subaru and HST data respectively. The values are given at the position of the quasar. For comparison, we also give the values from \citet{2010ApJ...711..201S}.
		$\kappa_{\mathrm{ext}}$ shows the median value, $\sigma_{\kappa}^{-}$ and $\sigma_{\kappa}^{+}$ correspond to the deviation from the 16th and the 84th quantiles respectively. For each method we indicate the Bayes Factor calculated with respect to the result from \citet{2010ApJ...711..201S}.}
	\label{tab:kappa_values}
\end{table}

In order to obtain the line of sight convergence in the field of \bquad, we combine the results from Section~\ref{sec:convergence_maps} and Section~\ref{sec:joint_convergence_distribution}. We construct $P(\kappawl | \data)$ by measuring $\kappawl$ at the position of \bquad\ in 300 and 250 Jackknife realisations of the convergence maps for the Subaru and the HST data respectively for all noise filtering techniques. 

We use the joint distributions $P(\kappaext | \kappawl)$ from the MS to translate $P(\kappawl | \data)$ to $P(\kappaext| \data)$, thereby accounting for shape noise, source redshift mismatch, and for the smaller scales possibly washed out during the spatial filtering of the weak lensing maps. 

The final $P(\kappaext| \data$ distributions are displayed in Fig.~\ref{fig:final_result}, together with the results from \citet{2010ApJ...711..201S} based on the weighted galaxy number counts. The values of $\kappaext$ at the position of \bquad\ and the associated error bars are given in Table~\ref{tab:kappa_values}. 

The Subaru data indicates lower convergence at the position of \bquad\ than the HST data. This is due to a lower resolution and sensitivity of the ground-based data owing to the smaller galaxy density and the larger smoothing necessary to filter the noise. The large convergence  comes from a localized area around the projected position of the quasar (see the middle and bottom panels in Figure~\ref{fig:kappa}), and is quickly decreasing with angular distance to it. Smoothing with a large kernel will thus result in a decrease of the signal.

\begin{figure}
	\centering
	\includegraphics[width=1\columnwidth]{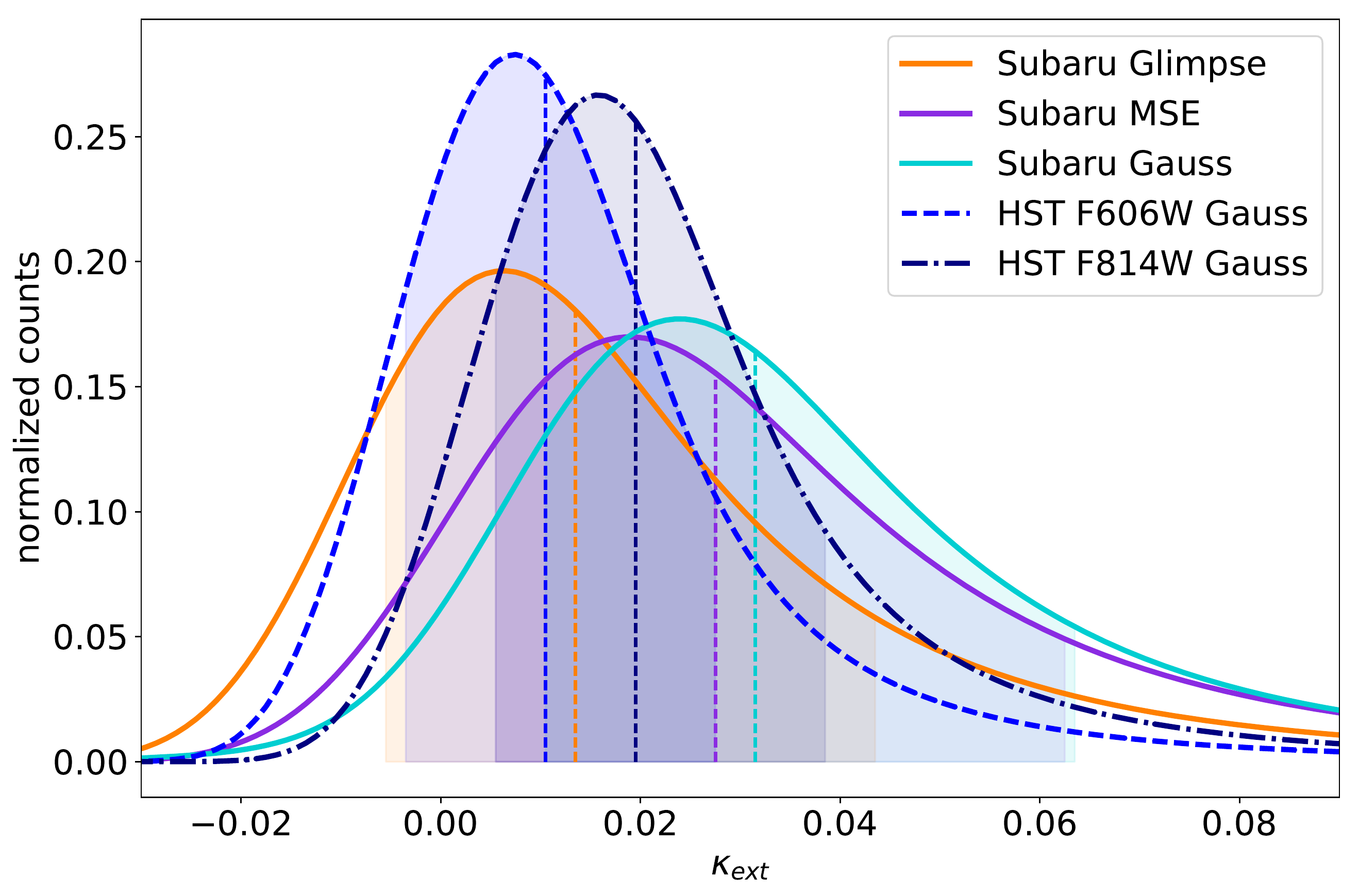}
	\caption{Illustration of the effect of spatial resolution of the weak lensing convergence $\kappawl$ on the PDF for the external convergence $\kappaext$ at the position of \bquad. The PDFs  are shown with the smoothing matching that of the Subaru data. The dashed blue, the dash-dotted black and the solid cyan lines show the Gaussian filtering with $\theta_{\mathrm{G}} = 0.75 \arcmin$ of the HST F606W, the HST F814W and the Subaru data respectively. The orange and the purple lines show the Glimpse and the MSE filtering of the Subaru data. The increased Gaussian smoothing kernel applied to the HST data ensures that the Subaru and HST mass maps have the same spatial resolution (see text).}
	\label{fig:HST_smoothed}
\end{figure}

To illustrate this, we apply filtering with a Gaussian kernel with $\theta_{\mathrm{G}} = 0.75 \arcmin$ to the HST data, i.e. the same Gaussian filtering that we use for the ground Subaru data, and carry out the Jackknife procedure. We then infer the $P(\kappa_{\mathrm{ext}} | \mathrm{\bf{d}})$ distributions at the position of the quasar. Figure~\ref{fig:HST_smoothed} shows the ensuing HST F606W and F814W distributions alongside the original Subaru ones. All the distributions are compatible with each other. Moreover, all reconstruction techniques produce compatible results, as discussed in Section~\ref{sec:comparison}. 
This also clearly illustrates that the ground-based data simply lack the spatial resolution necessary to see the convergence peak, i.e.  $\kappawl$, near the location of the quasar.

Note that the quantity relevant for cosmography and $H_0$ inference is the value of $\kappaext$ on scales comparable to the angular size of the whole \bquad\ system, i.e. about $5\,\arcsect$. This happens to match the spaxel size used for the HST data and shows that deep HST data are really needed for the present work. The spaxel size used for the Subaru data, imposed to us by the limited number density of background galaxies, is much larger ($14\,\arcsect$) and results in a bias in the $\kappaext$ values, as is illustrated in Fig.~\ref{fig:final_result}.

Further structures, smaller than 3-5\arcsec\ may be revealed by using higher resolution and deeper data than HST in combination with cosmological simulations also of higher spatial resolution. However, such variations, like mass gradients across the quasar images are already included explicitly in the lens model. In the event they are not, these variations would result in a broadening of our $\kappaext$ distributions rather than a systematic shift, because any small-scale and unknown $\kappaext$ contribution due to LoS can either be positive of negative, simply following cosmic variance. 

Since the spatial resolution of the mass maps obtained with the HST data greatly surpasses that of the Subaru data, we consider only the HST data for our final analysis. We marginalize over 6 $\kappaext$ distributions (not maps) since there is no reason to trust one filter more than the other and since we use 3 different methods for each filter. Our final value for $\kappaext$ evaluated in this way is  $\kappa = 0.11 ^{+0.06}_{-0.04}$, where the lower bound corresponds to the 16th percentile and the upper bound to the 84th percentile.

\section{Comparison with previous work}
\label{sec:comparison}

\begin{figure*}
	\centering
	\captionsetup[subfigure]{labelformat=empty}
	\subfloat{
		\includegraphics[width=1\textwidth]{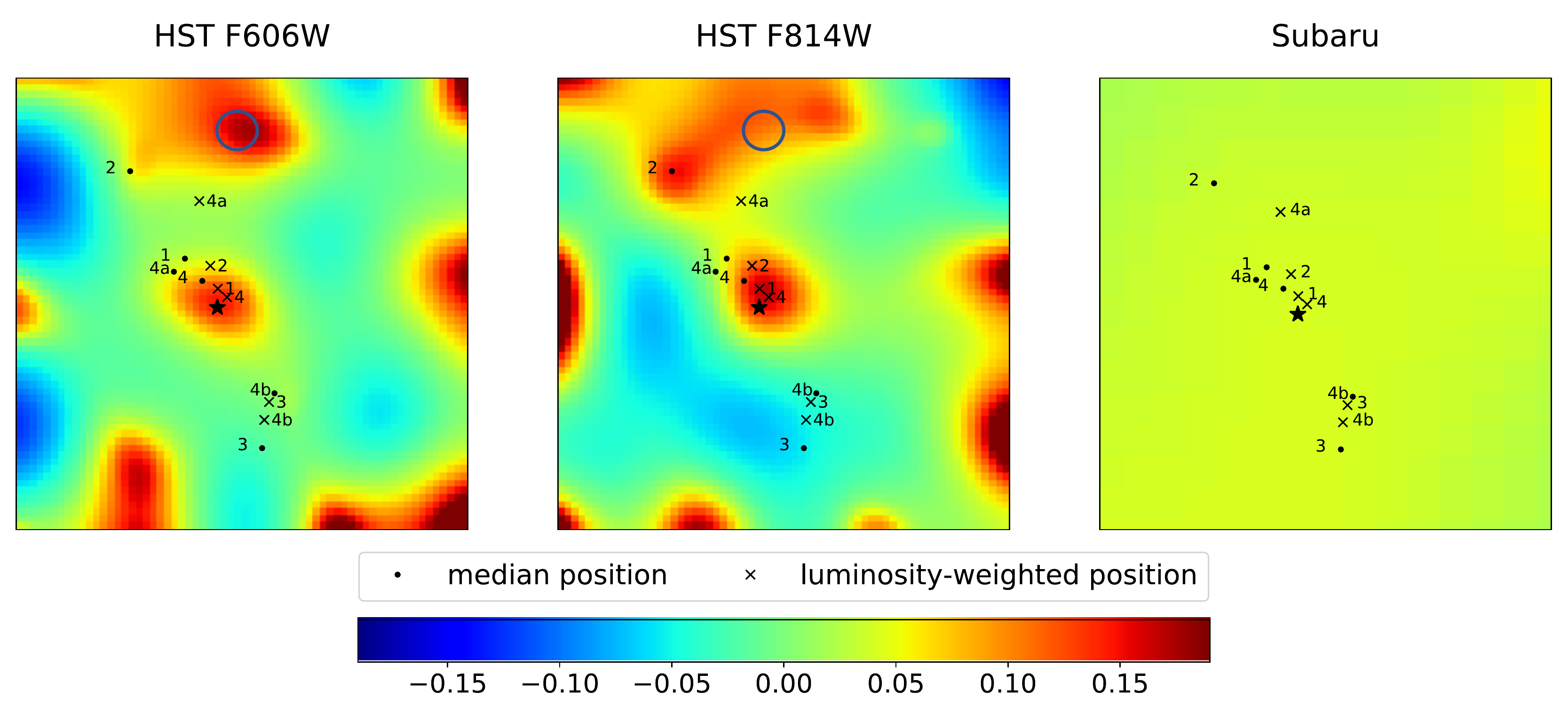}}
	\caption{The galaxy groups identified in \citet{2006ApJ...642...30F} are shown in overlay of the MSE filtering mass maps for the HST and Subaru data. The field of view is the same for the 3 datasets, i.e. $3.21\arcmin \times 3.21 \arcmin$. \bquad\ is indicated with a star symbol. The crosses and circles show the median and luminosity-weighted centroids of the groups shown in Table~1 of \citet{2006ApJ...642...30F}. The blue circle indicates a compact group seen in Fig.~\ref{fig:1608}, but not reported in \citet{2006ApJ...642...30F} (see text). }
	\label{fig:groups_kappa_map}
\end{figure*}

\begin{figure*}
	\centering
	\captionsetup[subfigure]{labelformat=empty}
	\subfloat{
		\includegraphics[width=0.7\textwidth]{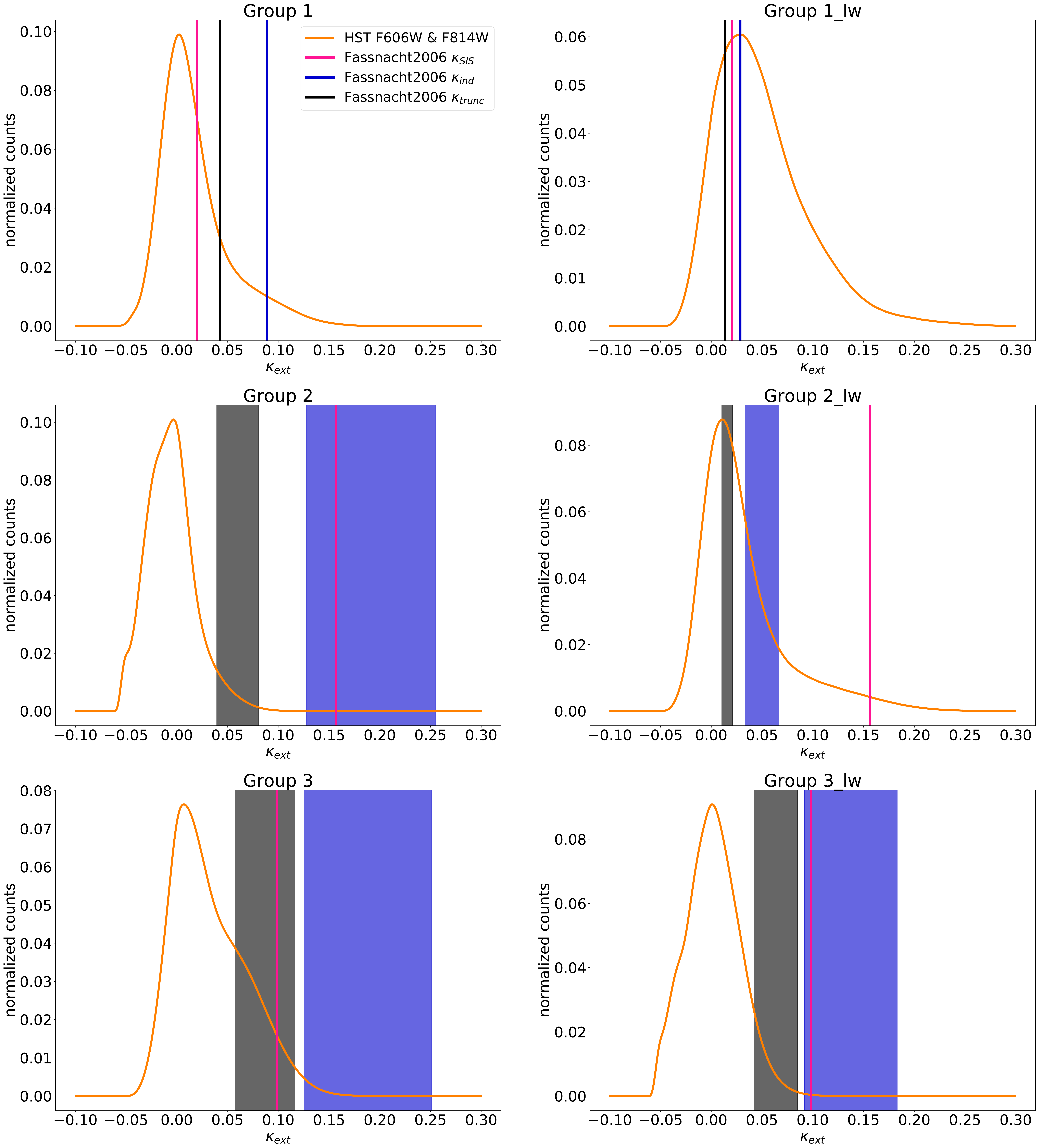}}
	\caption{The PDF's for the convergence at the position of the groups found in \citet{2006ApJ...642...30F}. The orange distributions show the weak lensing results, considering only the HST data (see end of Section~\ref{sec:results}). The vertical lines and the shaded regions depict the convergence values from \citet{2006ApJ...642...30F}, according to their Table~2. The pink line corresponds to their estimates of the convergence when modeling thew group with a single SIS halo. The black and blue lines give the result when modeling the groups as a collection of halos with and without a truncation radius, respectively. The shaded regions indicate the uncertainty from the estimates of the velocity dispersion for the individual galaxies, while the vertical lines denote the estimates without the quoted uncertainty. The left column corresponds to the group centroids determined as the median of the member positions, while the right column shows the estimate as the luminosity-weighted mean.}
	\label{fig:groups_pdf1}
\end{figure*}

\renewcommand{\thefigure}{\arabic{figure} (continued)}
\addtocounter{figure}{-1}
\begin{figure*}
\centering
\captionsetup[subfigure]{labelformat=empty}
\subfloat{
	\includegraphics[width=0.7\textwidth]{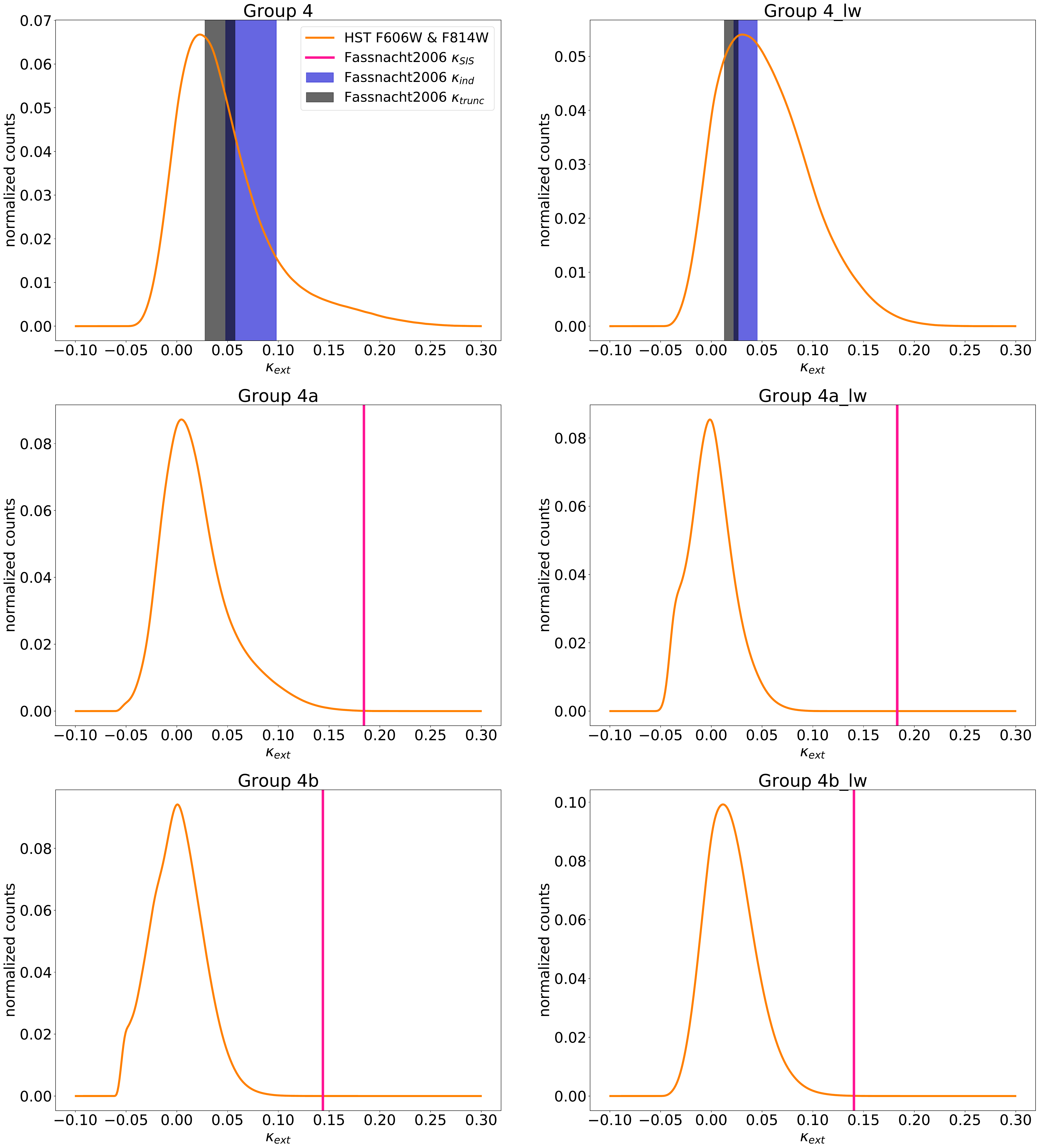}}
	\caption{}
	\label{fig:groups_pdf2}
\end{figure*}
\renewcommand{\thefigure}{\arabic{figure}}

The environment of \bquad\ was previously studied  with different methods, that we compare with our results. Using the number counts technique \citet{2011MNRAS.410.2167F} found that \bquad\ resides in an over-dense environment, being roughly twice as dense as control fields. \citet{2010ApJ...711..201S} used the result of \citet{2011MNRAS.410.2167F} to select lines of sight in the MS containing the same relative over-density, and extract the distribution of $\kappa_{\mathrm{ext}}$. The resulting distribution is shown in Figure~\ref{fig:final_result} and comparesd with our new weak lensing distributions. Although all distributions look well  compatible, we use the Bayesian formalism proposed in \citet{2006PhRvD..73f7302M} to  quantify this and test two hypotheses, $H_\mathrm{1}$ and $H_\mathrm{2}$:

\begin{enumerate}
\item $H_\mathrm{1}$: the two distributions for $\kappaext$, from weak lensing and from number counts, can be consistently explained within one single underlying cosmology under the assumption that no unknown systematic bias affects one or the other $\kappaext$ measurement.

\item $H_\mathrm{2}$: the two distributions or one of them suffer from its/their own systematic error or are not drawn from the same cosmology or underlying physical model. 
\end{enumerate} 

The question is then to know whether the data favors the $H_\mathrm{1}$ or the $H_\mathrm{2}$ hypothesis, i.e. whether the data points towards the presence of unknown systematics that need to be explicitely accounted for, or not. For this purpose, we calculate the Bayes factor $F$, given by
\begin{equation}
F = \frac{ P(\kappaextwl, \kappaextnc | H_\mathrm{1}) }{ P(\kappaextwl,  \kappaextnc | H_\mathrm{2}) },
\end{equation}
where $\kappaextwl$ is the external convergence inferred from weak lensing and $\kappaextnc$ is the external convergence inferred from the number-count technique. Following the appendix of \citet{2013ApJ...766...70S}, this boils down to
\begin{equation}
F = \frac{ \langle \Lextwl \, \Lextnc \rangle }{ \langle \Lextwl \rangle  \langle \Lextnc \rangle },
\end{equation}
where $\Lextwl$ and $\Lextnc$ are the likelihoods of respectively $\kappaextwl$ and $\kappaextnc$. If $F>1$, the data is in favor of $H_\mathrm{1}$ hypothesis. For reference, two one-dimensional Gaussian likelihoods have a Bayes factor of $F = 1$ if they overlap within $2\sigma$, and $F\approx 3.6$ if the two distributions overlap within $1\sigma$.

The computed Bayes factors with respect to the result from \citet{2010ApJ...711..201S} for all convergence reconstruction techniques for both ground-based and space-based data are shown in Table~\ref{tab:kappa_values}. The HST data yields very good agreement with the number counts technique, while the Subaru data is slightly discrepant. This results from the fact that the ground-based data cannot probe spatial scales as small as space-based data. All three reconstruction and filtering techniques are inter-compatible, with comparable errors and corresponding Bayes factors being $F>10$. We therefore conclude that there is no hidden systematic that makes any pair of dataset incompatible with each other.

\section{Convergence of groups in the vicinity of \bquad}
\label{groups}

\citet{2006ApJ...642...30F} examine the neighborhood of \bquad. They find four groups of galaxies within a projected radius of $1\arcmin$ from the lensed system. The fourth of these groups may be a merger event or a filamentary structure, so that the authors divide it into two subgroups, based on the redshifts of the members. For each of these groups and subgroups, \citet{2006ApJ...642...30F} provide a convergence value based on analytical modeling or their halo, which we compare here to our non-parametric reconstructions. 

In \citet{2006ApJ...642...30F}, the centroids of the groups are estimated in two ways. First using the median of the member positions and, second, by using their luminosity-weighted mean. The locations of the groups from the two methods are shown in Fig.~\ref{fig:groups_kappa_map} in overlay of our HST weak lensing maps and show a clear correlation between the higher convergence and the presence of the groups. 

\citet{2006ApJ...642...30F} determine the convergence of the groups with two different approaches. The first assumes that all members of a given group share the same halo, modelled as a Singular isothermal sphere (SIS) located at the mean position of the group or at the luminosity-weighted position. The second approach considers each group as a collection of individual halos, with the final convergence being the sum of contributions from all members. Each separate halo is modelled as an SIS, with or without truncation radius \citep[see Table~2 of][]{2006ApJ...642...30F}. These are compared to our HST weak lensing convergence values in Fig.~\ref{fig:groups_pdf1}. In all cases, the halo with a truncation radius are prefered by the weak lensing data, especially in dense environments \citep[e.g.][]{2007A&A...461..881L, 2013ApJ...774..124E}.

\citet{2006ApJ...642...30F} discuss the presence of a potential cluster at redshift $z = 0.52$, labeled here as Group 4. In order to test the hypothesis of a merger event rather than a single cluster, they divide it into two subgroups, Group 4a and Group 4b. Our weak lensing results are consistent with zero convergence at the position of these subgroups, disfavoring the two-group hypothesis. However, our convergence distributions at the position of Group 4 hint towards one single-halo and our convergence estimates are consistent with that of \citet{2006ApJ...642...30F} in the single-halo hypothesis. 

Finally, we also note a rich and compact group of galaxies at the very top of the HST field in Fig.~\ref{fig:1608}, that is well detected in our weak lensing maps but not at all in \citet{2006ApJ...642...30F}. This may be due to the compactness of the cluster or group, preventing these authors to place spectroscopic slits in a sufficiently compact configuration to measure redshifts. This compact group is indicated with a blue circle in Fig.~\ref{fig:groups_kappa_map}.

\section{Summary}
\label{sec:summary}

We analysed the environment and the line of sight of the quadruply lensed quasar \bquad, which is one of the lenses in the H0LiCOW sample. From galaxy number counts, \bquad\ is already known to reside in a highly over-dense environment, substantially impacting the cosmological analysis from time-delay cosmography. We provided a complementary and independent measurement of the external convergence at the position of the system using weak gravitational lensing. 

For this purpose we used deep Subaru Suprime-Cam observations and HST data in the F606W and F814W filters. The source galaxy density is 14 gal/arcmin$^2$ and 220 gal/arcmin$^2$ for the Subaru and HST data respectively. We adopted three different methods to reconstruct the convergence maps: Kaiser-Squires inversion of the inpainted shear field with either Gaussian or multi-scale entropy noise filtering, and the sparsity based $\tt{Glipmse}$ technique. The inpainting was used to remove the noise coming from the unevenly sampled shear data and the gaps in the field, propagating to the convergence field. However, the inpainted data themselves were not used in the final analysis. We adopted two different Gaussian smoothings for the Subaru and the HST, i.e. $\theta_{\mathrm{G}} = 0.75\arcmin$ and $\theta_{\mathrm{G}} = 0.15\arcmin$ kernels respectively. 

We find that the mass maps based on Kaiser-Squires inversion and $\tt{Glipmse}$ algorithm are qualitatively very similar. Without surprise, reconstructions with the ground-based data present less small-scale features as with the space-based data. This is due to the need of applying larger smoothing kernels for the ground-based data than for space-based data in order to ensure similar signal-to-noise for the shears maps. 

The convergence obtained from a weak lensing reconstruction is expected to differ from the external convergence due to, e.g, shape noise in the weak lensing shear, different source redshift distributions and different spatial smoothing scales. We thus employed lensing simulations based on the MS to estimate the joint distribution of the external convergence and the weak lensing convergence for the different source galaxy catalogs and convergence reconstruction methods. From that we computed the posterior probability distribution of the external convergence given the weak lensing data. 

Due to their lower resolution and sensitivity, we do not use the Subaru data for our final results. Based solely on the HST data, we find that the environment and the line of sight of \bquad\ is effectively over-dense. Our measurements are compatible with the previous analysis using number counts technique by \citet{2011MNRAS.410.2167F} and \citet{2010ApJ...711..201S}. However, our methodology is independent and complementary, as it probes directly the distributions of dark and luminous matter in a non-parametric and is not model dependent.  Owing to the agreement between the weak lensing and number-counts results by \citet{2010ApJ...711..201S}, there is no need for any additional correction to the $\kappaext$ value adopted in previous work based on \bquad.

Finally, we investigate the presence of groups found in \bquad\ field by \citet{2006ApJ...642...30F}. Our HST mass maps show  correlation between the positions of the groups and the peaks in the convergence map. We also find that the convergence value found from the weak lensing analysis agree better with truncated halo models for the groups than with non-truncated halos.

\section*{Acknowledgments}
This work is supported by the Swiss National Science Foundation (SNSF). This project has received funding from the European Research Council (ERC) under the European Union's Horizon 2020 research and innovation program (COSMICLENS: grant agreement No 787886). O.T. acknowledges support from the Swiss Society for Astronomy and Astrophysics (SSAA), and thanks Sandrine Pires and Yves Revaz for fruitful discussions and suggestions. This work was supported by World Premier International Research Center Initiative (WPI Initiative), MEXT, Japan.  K.C.W. is supported in part by an EACOA Fellowship awarded by the East Asia Core Observatories Association, which consists of the Academia Sinica Institute of Astronomy and Astrophysics, the National Astronomical Observatory of Japan, the National Astronomical Observatories of the Chinese Academy of Sciences, and the Korea Astronomy and Space Science Institute. S.H.S.~thanks the Max Planck Society for support through the Max Planck Research Group. S.H. acknowledges support by the DFG cluster of excellence \lq{}Origin and Structure of the Universe\rq{} (\href{http://www.universe-cluster.de}{\texttt{www.universe-cluster.de}}). C.E.R. and C.D.F. acknowledge support for this work from the National Science Foundation under Grant No. AST-1312329, and C.D.F. further acknowledges support from the NSF under Grant No. AST-1715611.

\bibliography{b1608wl}
\bibliographystyle{mnras}

\bsp	
\label{lastpage}
\end{document}